\documentclass[aps,twocolumn,showpacs,preprintnumbers,amsmath,amssymb]{revtex4}
\usepackage{graphicx}
\usepackage{dcolumn}
\usepackage{color}
\usepackage{tabularx}
\usepackage[utf8]{inputenc}
\usepackage{amssymb}
\usepackage{bm}

\def\bea {\begin{eqnarray}}
\def\eea {\end{eqnarray}}

\def\be {\begin{equation}}
\def\ee {\end{equation}}
\def\nn {\nonumber}

\begin{document}

\title{Thermalization of dense hadronic matter in Au~+~Au collisions at energies available at the 
Facility for Antiproton and Ion Research}
\author{Somnath De$^1$, Sudipan De$^2$, and Subhasis Chattopadhyay$^3$}

\affiliation{$^1$ Institute of Physics, Bhubaneswar, Odisha, India}
\affiliation{$^2$ Universidade de S\~{a}o Paulo, S\~{a}o Paulo, Brasil}
\affiliation{$^3$ Variable Energy Cyclotron Centre, 1/AF, Bidhan Nagar, Kolkata, India}

\medskip

\begin{abstract}
The conditions of local thermodynamic equilibrium of baryons (non-strange, strange) and mesons (strange) 
are presented for central Au + Au collisions at FAIR energies using the microscopic transport model UrQMD. The net particle 
density, longitudinal-to-transverse pressure anisotropy and inverse slope parameters of the energy spectra 
of non-strange and strange hadrons are calculated inside a cell in the central region 
within rapidity window $|y| < 1.0$ at different 
time steps after the collisions. We observed that the strangeness content is dominated by baryons at all energies, however 
contribution from mesons become significant at higher energies. The time scale obtained from local pressure (momentum) 
isotropization and thermalization of energy spectra are nearly equal and found to decrease with increase in laboratory energy. 
The equilibrium thermodynamic properties of the system are obtained with statistical thermal model. The time evolution of the
entropy densities at FAIR energies are found very similar with the ideal hydrodynamic behaviour at top RHIC energy.
\end{abstract}

\maketitle


\section{Introduction}
The motivation of the relativistic heavy ion collider experiments is to explore the properties of 
strongly interacting matter (partonic or hadronic) at the finite temperature and/or density.
The current heavy ion research facilities e.g; Relativistic Heavy Ion Collider (RHIC) and Large Hadron Collider (LHC) are 
focused in unveiling the properties of deconfined quark-gluon matter created at the extreme 
temperature and almost vanishing net baryon density~\cite{RHIC-Exp,LHC-Exp}. At this regime the lattice quantum chromodynamics (lQCD) simulations 
have reported a crossover from hadronic to partonic phase and the existence of critical point 
where the first order phase transition line terminates~\cite{lqcd}. Thus RHIC has initiated the beam energy 
scan program to find the location of critical point in the QCD phase diagram 
(temperature (T)- baryo-chemical potential (\textrm{$\mu_B$}) plane)~\cite{BES}.

In contrary to the above experiments, the future Compressed Baryonic Matter (CBM) experiment at FAIR / GSI laboratory is aimed to explore another facet of QCD phase 
diagram; at high baryon density ($\sim 7-8$ times ground state nuclear matter density) and moderate temperature~\cite{CBM-exp1,CBM-exp2}. 
The experiment would be playing a very significant role in the scientific quest of understanding the behaviour of QCD at high
density regime. The facility is being designed to collide various species of heavy ions at fixed target mode with the anticipated beam 
energies 5-45 GeV/nucleon. The diagnostic probes of the matter created in the collisions include; (i) Short lived vector mesons ($\rho$, $\omega$) decaying to 
dilepton pairs, (ii) Production of multi-strange hyperons ($\Xi$, $\Omega$) (iii) Dissociation of Charmonium (J/$\Psi$) and 
charmed hadron ($D$, $\Lambda_c$) states, etc. The existence of first order phase transition from hadronic to 
partonic matter and restoration of chiral symmetry at the large \textrm{$\mu_B$} is expected to be found from the FAIR energy scan program~\cite{P.Senger}. 
Earlier experiments such as; RHIC-AGS and CERN-SPS were aimed to explore above features through 
the measurement of bulk observables like; flow and momentum spectra of hadrons. However their efforts were constrained due to 
limited beam luminosity. In recent years a similar research program (NICA) at JINR-Dubna has been proposed to explore phases of 
nuclear matter at high baryon density~\cite{Dubna}. 
But the CBM experiment would be more efficient for the detection of bulk and rare probes, with the availability of 
high intensity ion beams~\cite{CBM-expt-PoS}.

In order to compute the dynamic evolution of the matter created in such collisions, we need macroscopic/microscopic 
models. The macroscopic models like; hydrodynamics rely upon the assumption of local 
thermal equilibrium of the created matter on a certain time scale. The actual thermalization criterion has seldom been tested. 
There are a few works, addressed the issue at higher collision energies in the framework of perturbative QCD~\cite{Baier} or color-glass condensate theory~\cite{Gelis}.
On the other hand microscopic Monte Carlo models like: UrQMD~\cite{urqmd-model}, HSD~\cite{hsd}, AMPT~\cite{ampt} work on the postulated interaction among it's constituents 
(parton, hadron, or string) and does not require any assumption of local thermal equilibrium. Therefore, it is very important to test weather the dense baryonic matter created in these 
collisions achieve a local thermal equilibrium or not. In particular, we have investigated the time-scale of local thermal 
equilibration of non-strange and strange baryons in an elementary volume in phase-space from the time evolution of 
longitudinal-to-transverse pressure anisotropy and slope of the energy spectrum. For this purpose, we have 
employed the microscopic, N-body transport model called Ultra-relativistic Quantum 
Molecular Dynamics (UrQMD). A comparison between the model and the data for central Pb+Pb collisions at different  
energies at CERN-SPS can be found in Ref.~\cite{NA49}. We considered the most central collisions of 
gold (Au) nuclei at four beam energies associated with the CBM experiment. The incident beam energy has obvious implication on the time-scale of equilibration, 
which can be found in the subsequent section. 

The organization of the paper goes as the following. In the next section, we briefly recapitulate about 
the microscopic transport model UrQMD and then discuss about the methodology of our analysis. In section 3, we show the results for
the time evolution of density, ratio of longitudinal-to-transverse pressure and inverse slope parameter of the energy spectra 
for non-strange baryons, strange baryons and strange mesons. In section 4, we have utilized the statistical thermal model to extract the 
post-equilibrium thermodynamic parameters e.g., temperature, chemical potentials and calculated the entropy density of the system. Finally we 
have summarized the findings in section 5.
\begin{figure*}
\begin{center}
\includegraphics[width= 7.5cm,clip= true]{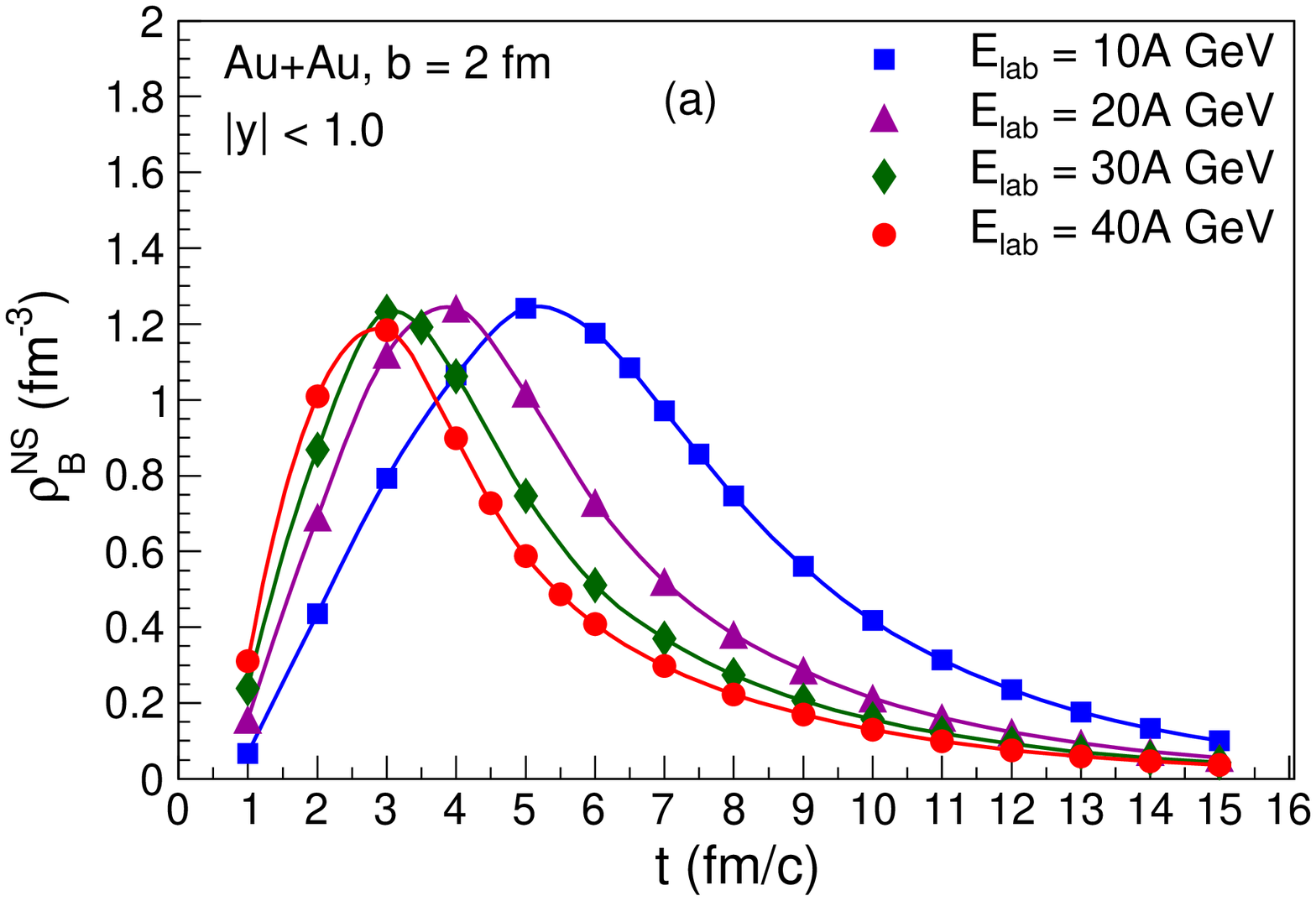}
\includegraphics[width= 7.5cm,clip= true]{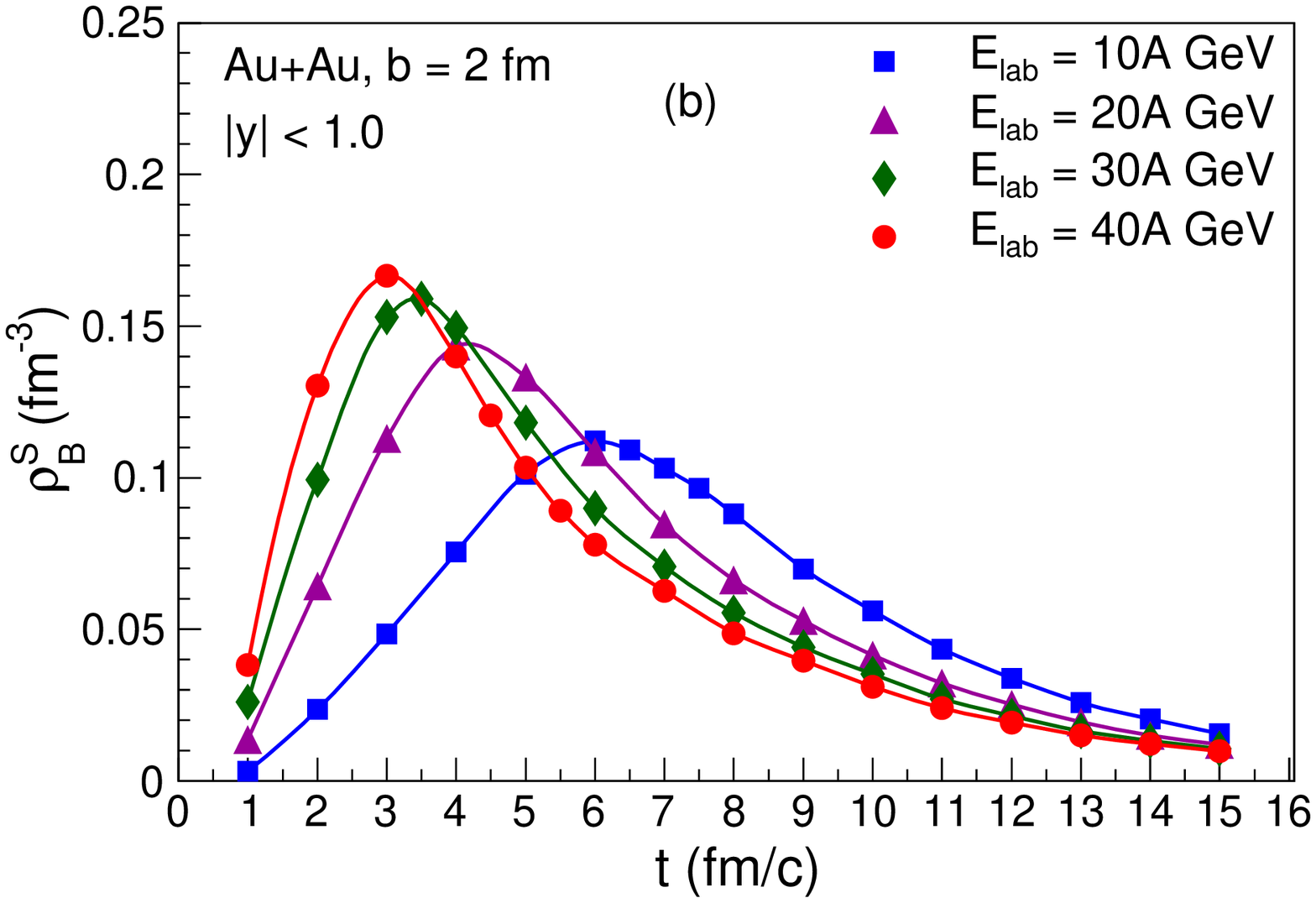}
\includegraphics[width= 7.5cm,clip= true]{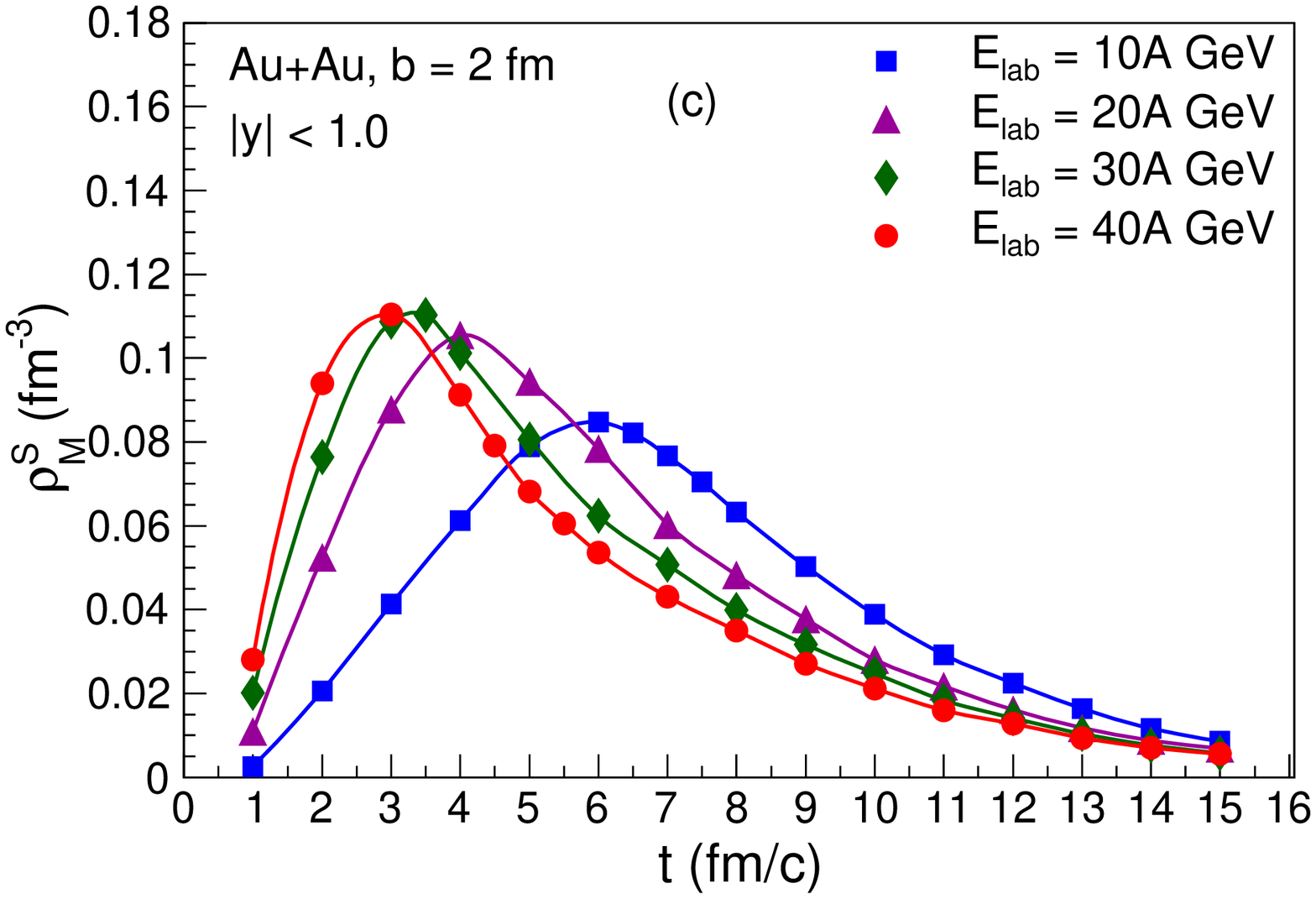}
\includegraphics[width= 7.5cm,clip= true]{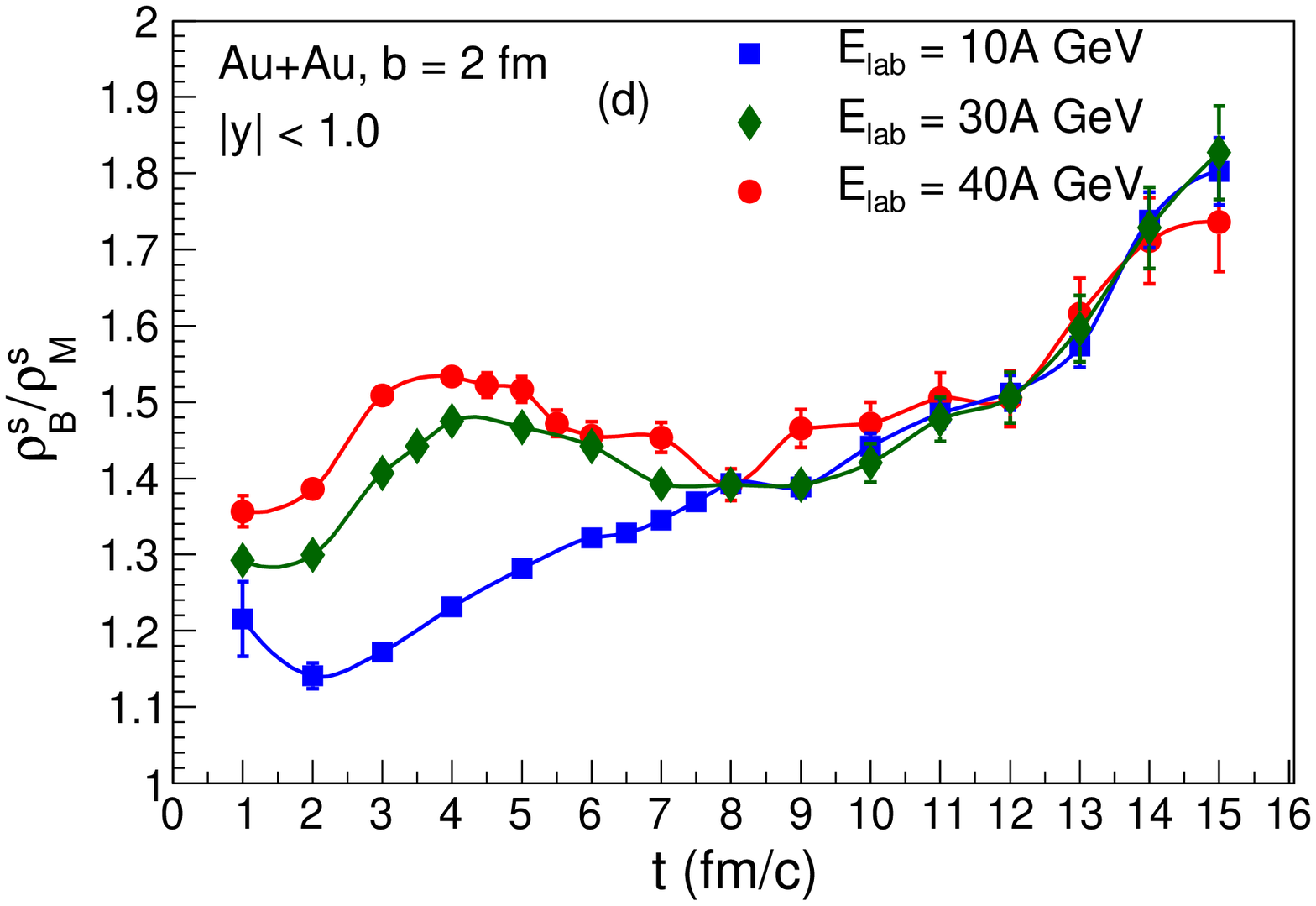}
\end{center}
\caption{(Color online)(Upper panel) Time evolution of net density of (a) non-strange baryons ($\rho_B^{NS}$), (b) 
strange baryons ($\rho_B^{S}$), (Lower panel) (c) kaons ($\rho_M^{S}$) and (d) net strange baryon to kaon ratio ($\rho_B^{S}$/ $\rho_M^{S}$) inside 
the central cell for Au~+~Au collisions (b = 2 fm) at the laboratory energies 10A, 20A, 30A, 40A GeV. The error bars are statistical only.}
\label{Fig-density}
\end{figure*}

\section{Methodology of the analysis}
The model UrQMD has been extensively used in recent years for describing heavy ion collisions of center of mass energy ranging from 
few GeV/ nucleon to few TeV/ nucleon~\cite{urqmd-model}.
We used the UrQMD-version:3.3p2 in default cascade mode without invoking any hydrodynamic evolution 
for the initial state. It includes 55 baryon species (up to mass 2.25 GeV) and 32 meson species (up to mass 
1.9 GeV) and their corresponding anti-particles and iso-spin projected states. Particle production in UrQMD occurs through
inelastic collisions, decay of meson, baryon resonances, and string fragmentation mechanism. At low energies 
($\rm{E_{lab}}<$ 4 GeV) hadronic interactions are based on two body or three body potential. However at high energies, hadron-hadron
collisions are performed stochastically in the spirit of cascade model~\cite{urqmd-cascade}. The total (elastic and inelastic) 
cross-sections of baryons and mesons are generally fitted to experimental proton-proton or proton-pion scattering data. For the resonant
baryon-meson or hyperon-baryon scattering where no experimental data are available, principle of detail balance or additive quark 
model have been used. The resonance scattering dominates the total cross-section at low beam momenta (upto $\rm{p_{lab}}\sim$ 2 GeV) 
however towards higher beam momenta string excitation has the largest contribution. The inelastic collisions and decays are responsible for 
changing the particle abundances of the system while the elastic collisions modify the momentum distribution of hadrons.
 
We have considered central collisions (impact parameter b = 2 fm) of Au nuclei 
at the laboratory energies ($\rm{E_{lab}}$) 10A, 20A, 30A, 40A GeV. For each energy we ran the simulation 
at different time steps ranging from 1fm/c to 15 fm/c. 6$\times 10^4$ events have been analyzed for each time step. 
The center of mass frame is chosen as the computational frame in our analysis.
We have considered a cell of dimension $2\times2\times2$ fm$^3$ about the origin of Au~+~Au system. The test volume has been 
chosen such that the effect of collective flow of the system on the observables will be minimum and at the same time the 
particle number should be large enough for reasonably small fluctuation in the observables.
Additionally a momentum rapidity cut $|y_{cm}|< 1.0$ has been imposed on the particles under consideration to ensure 
that the beam nucleon contribution does not come into account.
We have calculated the net particle density, different components of microscopic pressure for
non-strange baryons, strange baryons and mesons inside the cell. The non-strange baryons include Proton ($p$) and 
Neutron ($n$), the  strange baryons include Lambda ($\Lambda$), Sigma ($\Sigma$), Cascade ($\Xi$), Omega ($\Omega$) and 
the strange mesons include Kaons: $K^+$, $K^0$. All the higher mass resonances (baryon and meson) are allowed to decay.
We did not include $\Omega$ in the pressure calculation at $\rm{E_{lab}}$ = 10A and 20A GeV due to its limited statistics at lower energies. However we 
expect that inclusion of $\Omega$ does not modify any conclusion drawn in this work. 
We have also calculated the energy spectra (EdN/d$^3$p vs. E) of Protons and Lambdas inside the cell. Lastly the 
above quantities are statistically averaged over the number of events for each time step.
\begin{figure*}[t]
\begin{center}
\includegraphics[width= 7.5cm,clip= true]{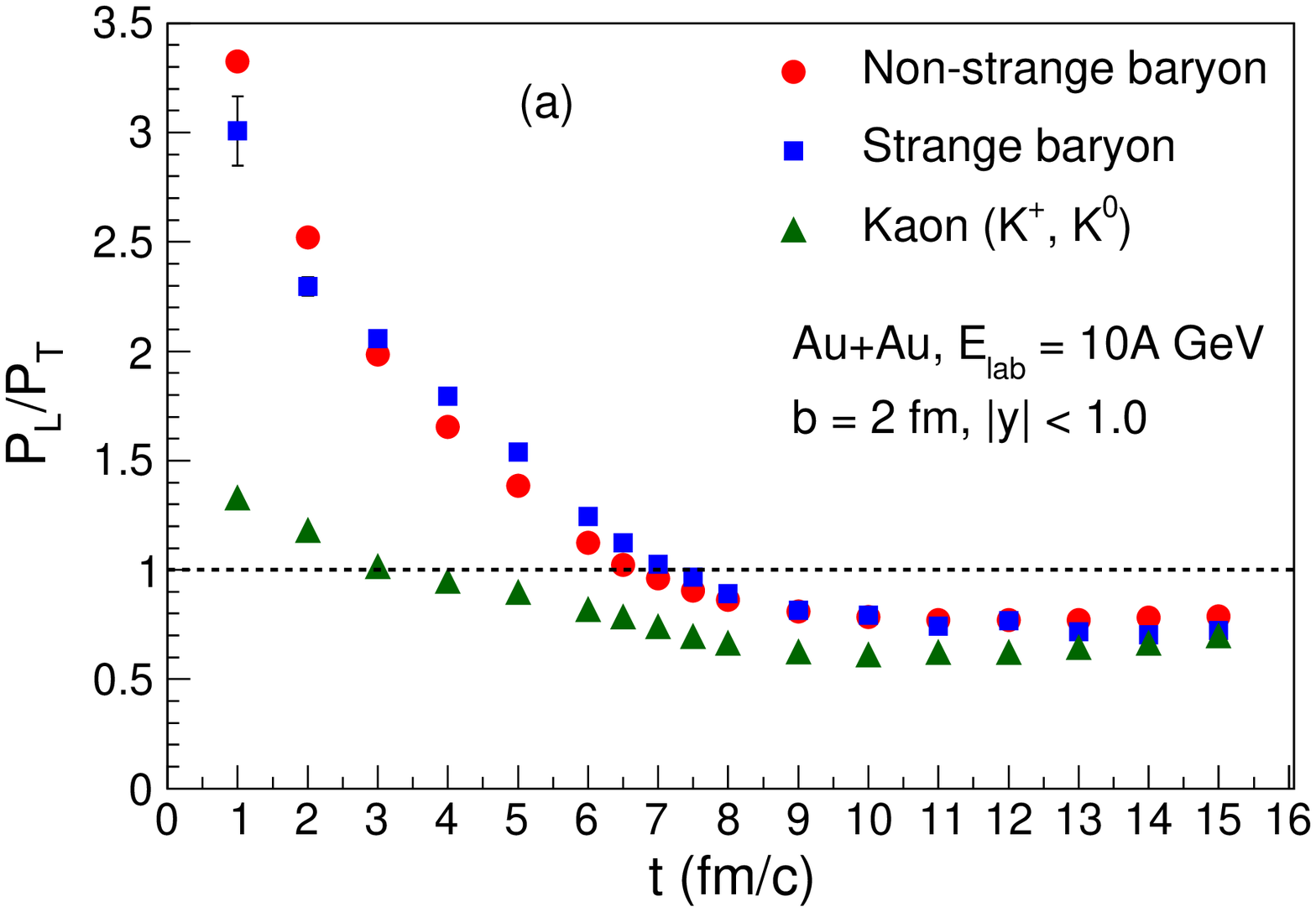}
\includegraphics[width= 7.5cm,clip= true]{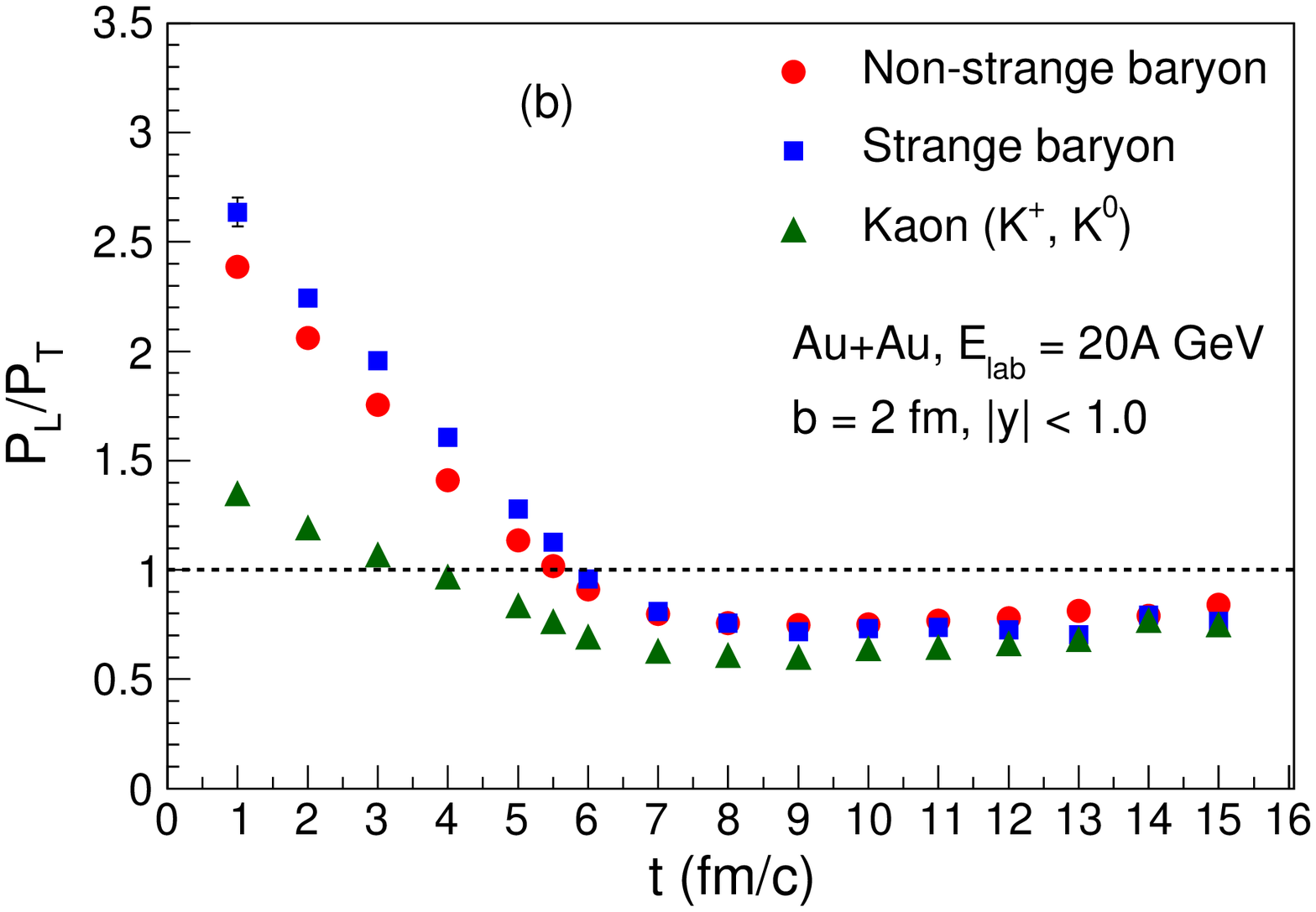}
\includegraphics[width= 7.5cm,clip= true]{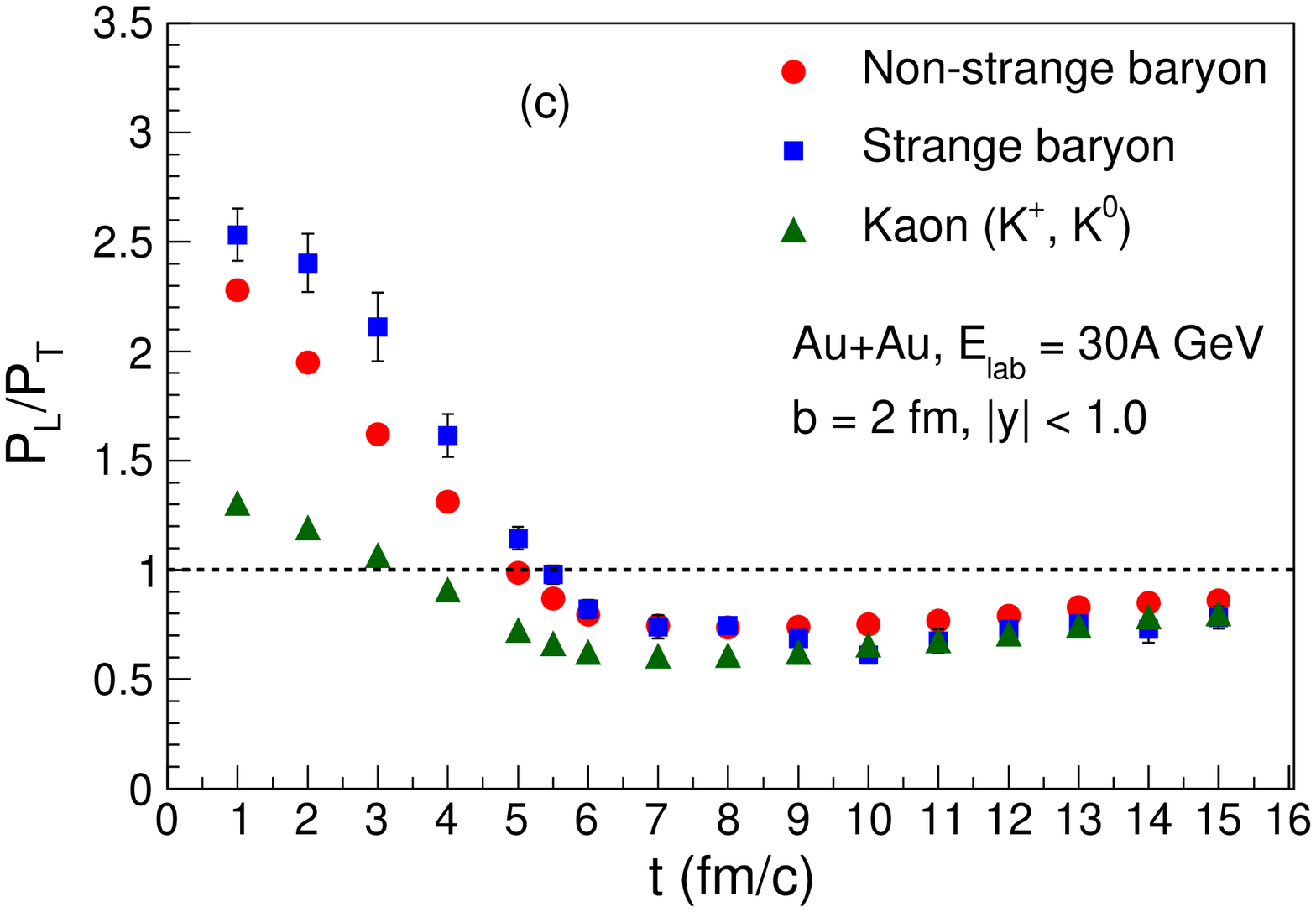}
\includegraphics[width= 7.5cm,clip= true]{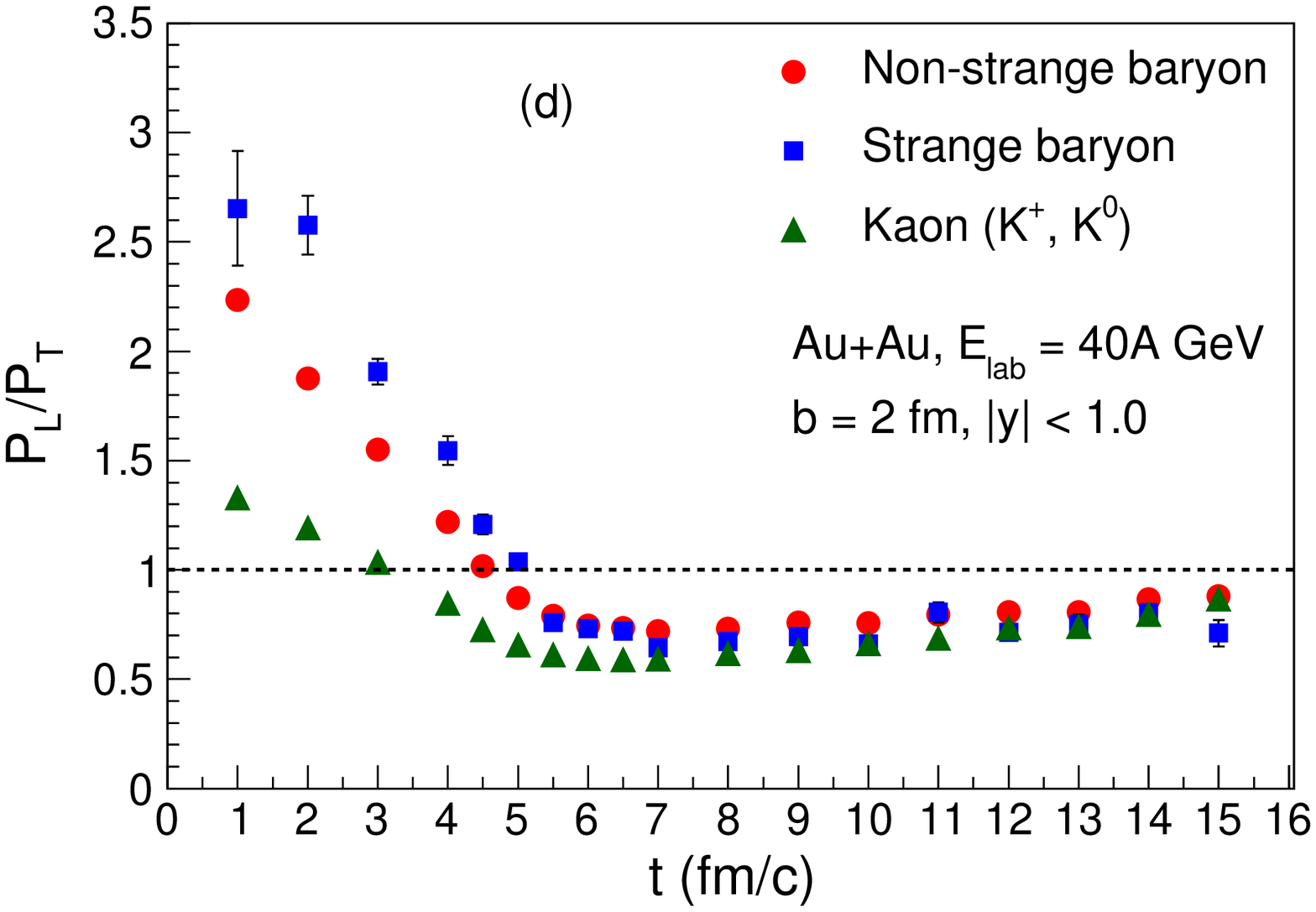}
\end{center}
\caption{(Color online) Time evolution of longitudinal-to-transverse pressure ratio ($P_L/ P_T$) of 
non-strange baryons, strange baryons, kaons inside the central cell for Au~+~Au collisions (b = 2 fm) at 
the laboratory energies (a) 10A, (b) 20A, (c) 30A, (d) 40A GeV. The error bars are statistical only.}
\label{Fig-pressure}
\end{figure*}
\section{Results}
\subsection{Time evolution of net particle density}
The time (t) is the elapsed time in center of 
mass frame. Time t  = 0 fm/c corresponds to the moment when two nuclei touch each other. The net particle density $\rho(t)$ is 
defined as the difference of particle density and anti-particle density. The evolution of net non-strange baryon 
density ($\rho_B^{NS}$), net strange baryon density ($\rho_B^{S}$), net kaon density ($\rho_M^{S}$), and 
net strange baryon to kaon ratio ($\rho_B^{S}$/ $\rho_M^{S}$) are depicted in Fig.~\ref{Fig-density} at $\rm{E_{lab}}$ = 10A, 20A, 30A, 40A GeV. 
The net particle density starts from a small value, reaches a maximum around t = 2R/($\gamma_{cm}v_{cm}$) when the two nuclei pass through 
each other and then falls down as the system expands. 
The generic feature has been found in agreement with earlier works~\cite{Dynamic-PRC,density-evolve}.
Here R is the radius of Au nucleus, $\gamma_{cm}$ and $v_{cm}$ are the Lorentz boost and velocity in center of mass frame. 
Thus we found the maximum matter density near 6 fm/c at 10A GeV and 3 fm/c at 40A GeV for all species. 
The production of non-strange baryons has been found almost similar for all the beam energies, but the strange baryon and meson production becomes 
larger with increasing beam energy. This is probably because of the fact that string excitation mechanism has 
major contribution to strangeness production at higher energies. The peak of baryonic (non-strange and strange) matter density
has been found at 40A GeV, which is about 7--8 times the ground state nuclear matter density. The time evolution of strange baryon 
to meson ratio has clearly shown the net strangeness content of the created matter is dominated by baryons for all the beam energies. 
The ratio has been found to grow with time because the kaons and Lambdas are produced through same strong interaction. 
However the kaons have suffered less scatterings in the medium due to its small interaction cross-section with other hadrons~\cite{Kaon-cross}, thus 
escapes the reaction volume quickly. The production of kaons is larger at higher beam energies can be seen from the non-monotonus 
behaviour of the ratio at smaller times.
\begin{figure*}[t]
\begin{center}
\includegraphics[width= 7.5cm,clip= true]{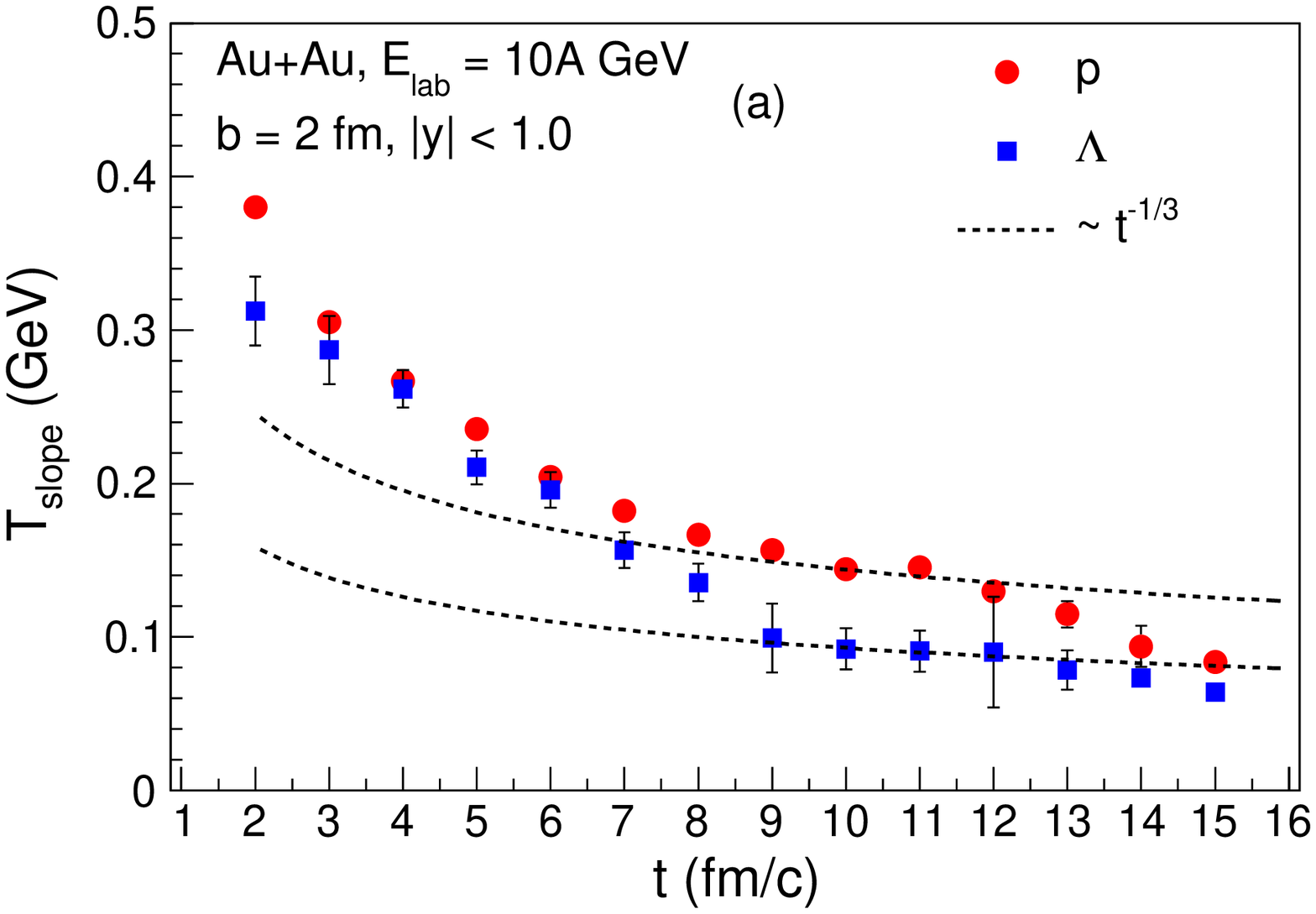}
\includegraphics[width= 7.5cm,clip= true]{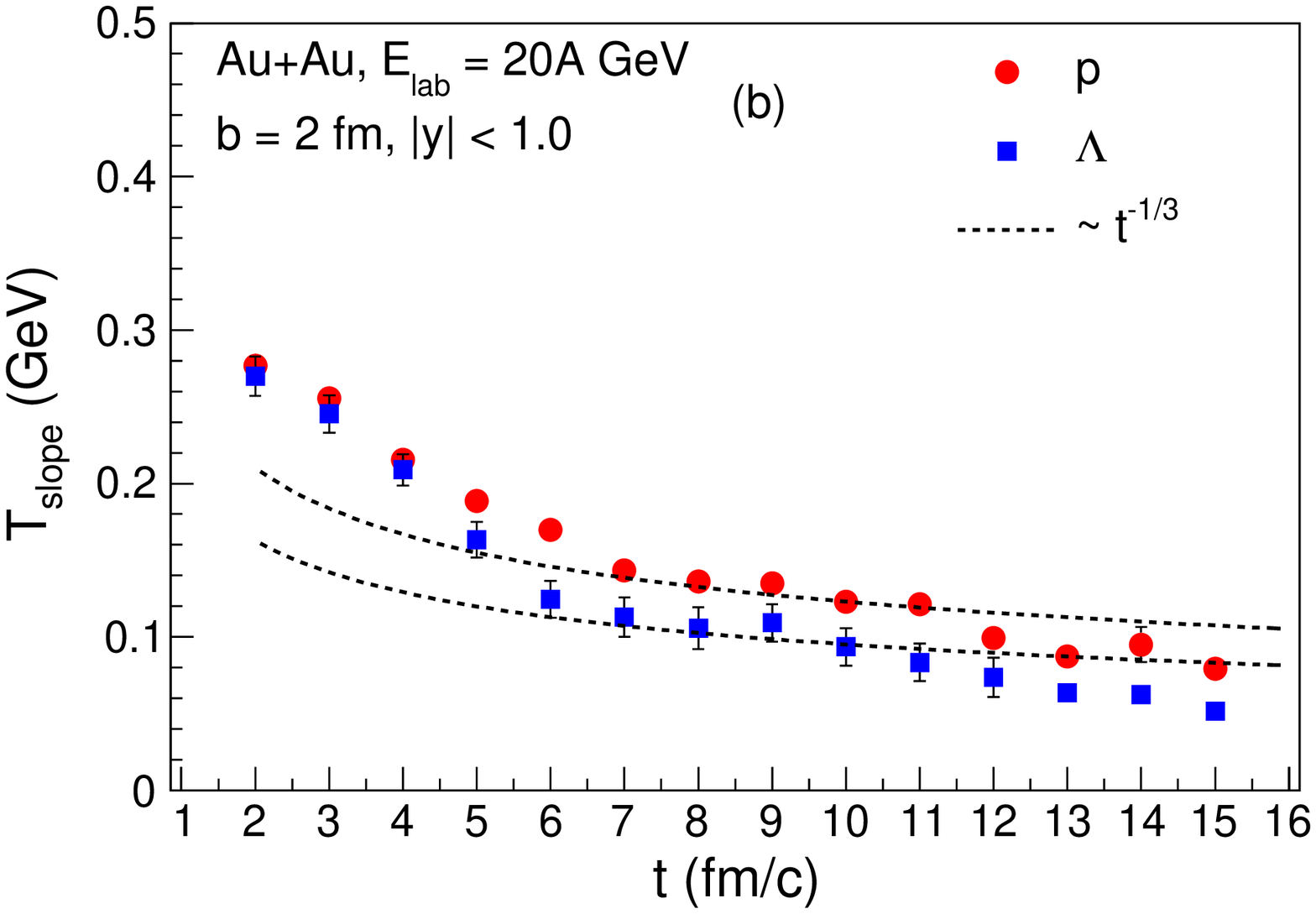}
\includegraphics[width= 7.5cm,clip= true]{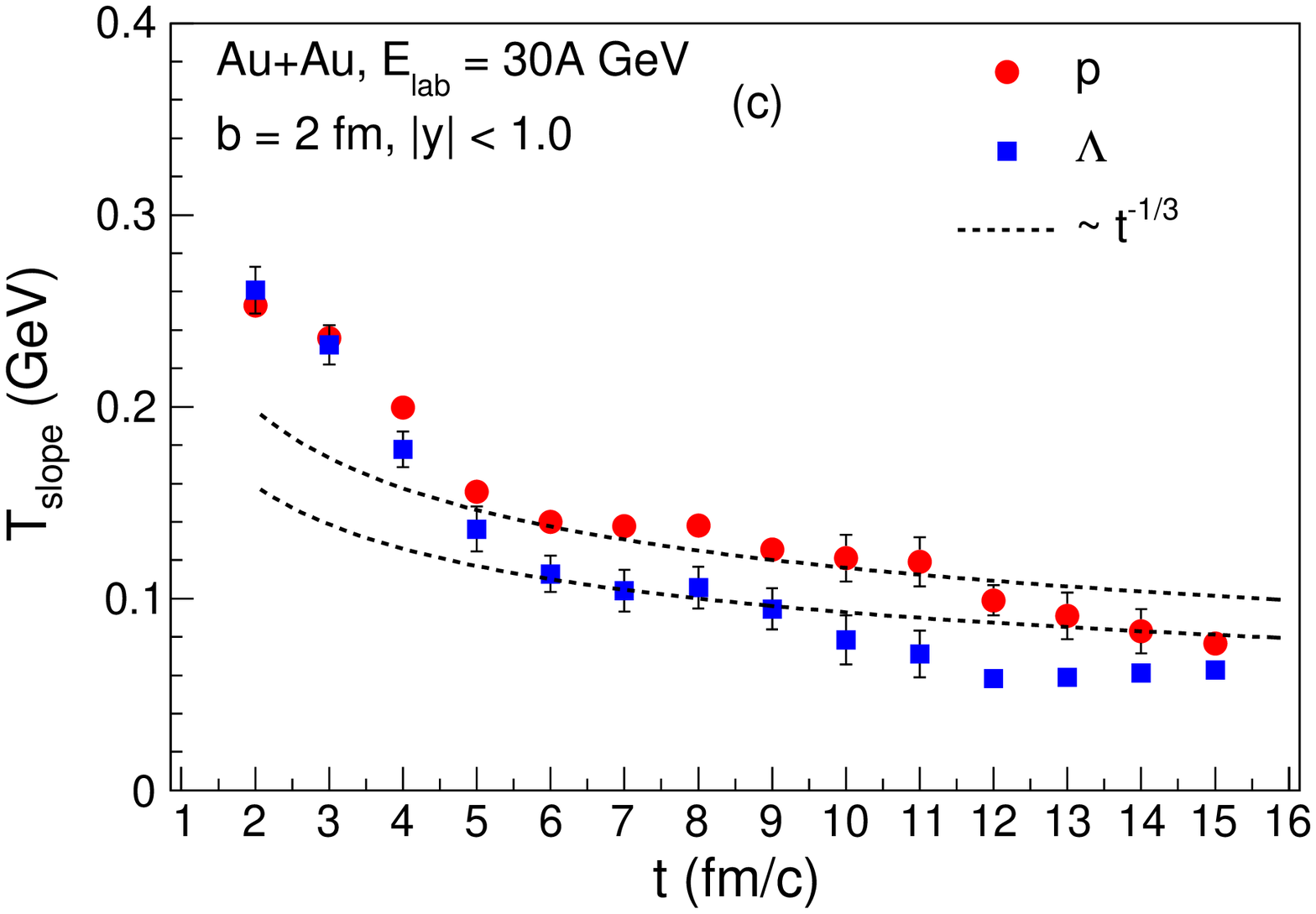}
\includegraphics[width= 7.5cm,clip= true]{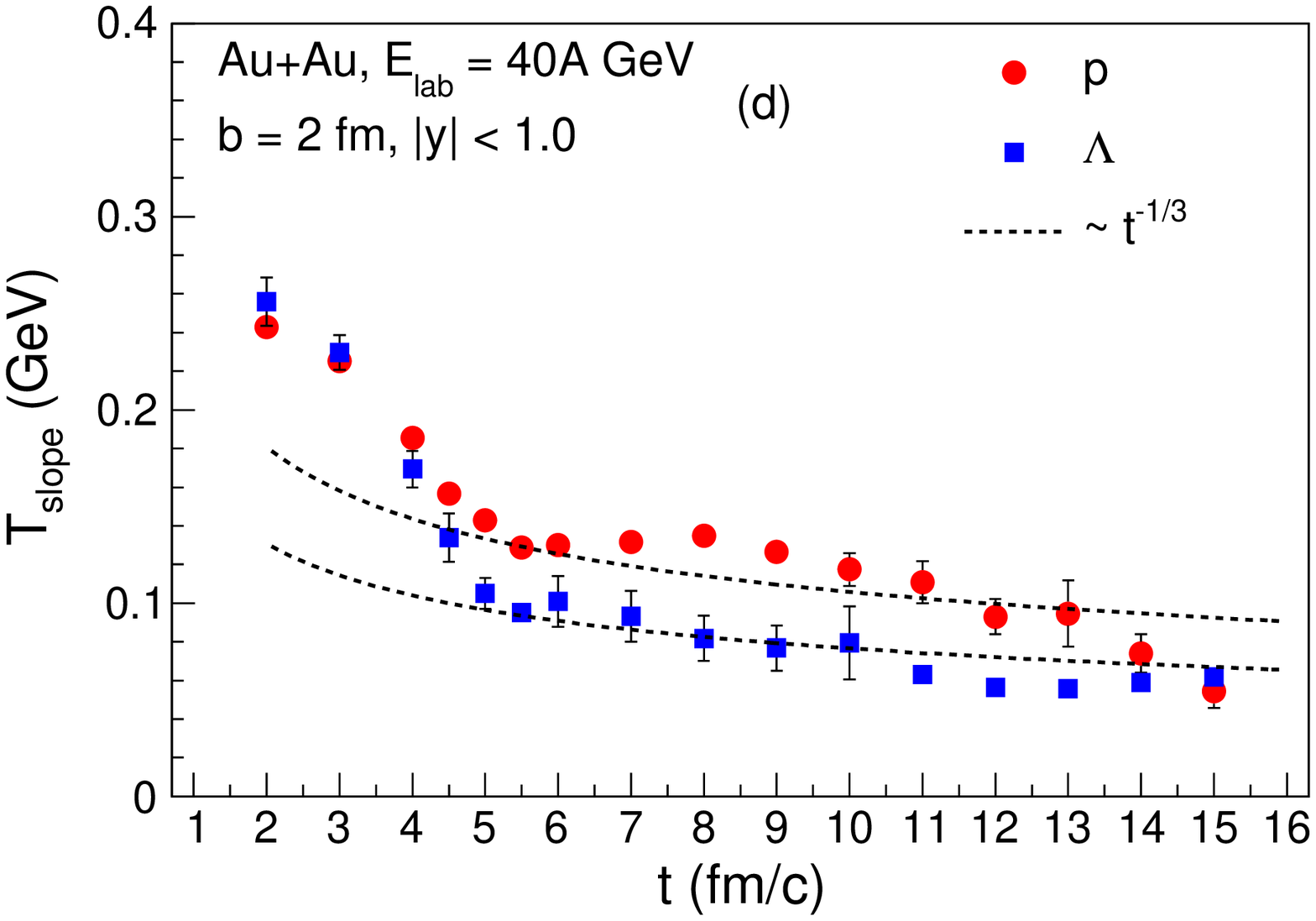}
\end{center}
\caption{(Color online) Time evolution of the inverse slope parameter ($T_{slope}$) of the energy spectra of Proton and Lambda 
inside the central cell for Au+Au collisions (b = 2 fm) at the laboratory energies (a) 10A, (b) 20A, (c) 30A, (d) 40A GeV. 
The error bars are statistical only.}
\label{fig-tslope}
\end{figure*}
\subsection{Isotropization of pressure components of baryons and mesons}
We have studied the isotropization of different components of microscopic pressure of non-strange baryons, strange baryons and 
kaons for an expanding system. The pressure components are highly anisotropic immediately after the collision. Thermal equilibrium is 
established in the cell when they have become nearly isotropic. Different components of microscopic pressure are calculated 
in UrQMD using ideal gas ansatz~\cite{UrQMD-1}:
\be
P_{(x,y,z)} = \sum_i \frac{p^2_{i(x,y,z)}}{3V(p^2_i+m^2_i)^{\frac{1}{2}}},
\ee
where, $p_i$ is the momentum, $m_i$ is the mass of i'th hadron and V is the volume of the cell under consideration. The longitudinal 
and transverse components of pressure for an ensemble of hadrons are defined as:
\be
P_L = \langle P_z \rangle;\hspace*{0.5cm} P_T = \frac{1}{2}(\langle P_x \rangle + \langle P_y\rangle),
\ee
here the $\langle ..\rangle$ corresponds to the statistical average over the number of events. The time evolution of the 
longitudinal-to-transverse pressure ratio ($P_L/P_T$) for the above mentioned hadron species are shown in 
Fig.~\ref{Fig-pressure} at the four beam energies.

The $P_L/ P_T$ ratio of baryons (non-strange and strange) starts from a large value at initial times, ultimately 
settles down to a value close to 1.0.  This reflects the longitudinal($z$) and transverse ($x,y$) momentum distribution of baryons 
are highly anisotropic at initial times. Successive elastic scatterings in the medium have made their momentum distribution nearly 
isotropic. We found that the ratio $P_L/ P_T$ becomes 1.0 around 6.5 fm for non-strange baryons 
and 7 fm for strange baryons at $\rm{E_{lab}}$ = 10A GeV. However the system further evolves and the ratio reaches a constant value $\sim$0.8 for 
t$\geq$ 9 fm/c. At this point we may say that the baryonic matter achieves a thermal equilibrium. Earlier work at AGS energy 
had also found similar time scale~\cite{UrQMD-1}. 
The deviation of $P_L/ P_T$ from unity after-equilibrium, is possibly arising due to finite shear viscosity of the hadronic 
matter~\cite{Bass-viscosity}. The ratio is more closer to unity as the system approaches towards ideal fluid limit. This has been 
shown by a recent study on the pressure isotropization in quark-gluon plasma for Au~+~Au coliisions at top RHIC energy~\cite{Greco}.
For other beam energies the $P_L/ P_T$ ratio of baryons has become unity much 
earlier, and it achieves a constant value $\sim$0.8--0.7 for t $\geq$ 8 fm/c at 20A GeV, for t $\geq$ 7 fm/c at 30A GeV and 
for t $\geq$ 6fm/c at 40A GeV. On close inspection of Fig.~\ref{Fig-pressure}, we found the pressure isotropization of non-strange baryons happen little earlier $\Delta t\sim$0.5 fm/c than strange 
baryons. The observation can be found in concurrence with an earlier UrQMD based calculation~\cite{strange-urqmd}, which has shown the average 
freeze-out time of nucleons is smaller than the strange baryons ($\Lambda$, $\Xi$). 
The $P_L/ P_T$ ratio of kaons approaches to 1.0 at early times $t\sim$3 fm/c and then 
becomes almost constant $\sim$ 0.6--0.7 at the same time as the baryons.
The initial longitudinal-to-transverse pressure (momentum) anisotropy of kaons has been found smaller than baryons, 
could be because of the facts that kaons have only one constituent quark from the original excited hadron and 
suffer less resonant scatterings than baryons in medium. Thus we found the pressure isotropization time 
of baryons and mesons reduces by about 3 fm/c from $\rm{E_{lab}}$ = 10A GeV to $\rm{E_{lab}}$ = 40A GeV. 
\begin{figure*}[t]
\begin{center}
\includegraphics[width= 7.5cm,clip= true]{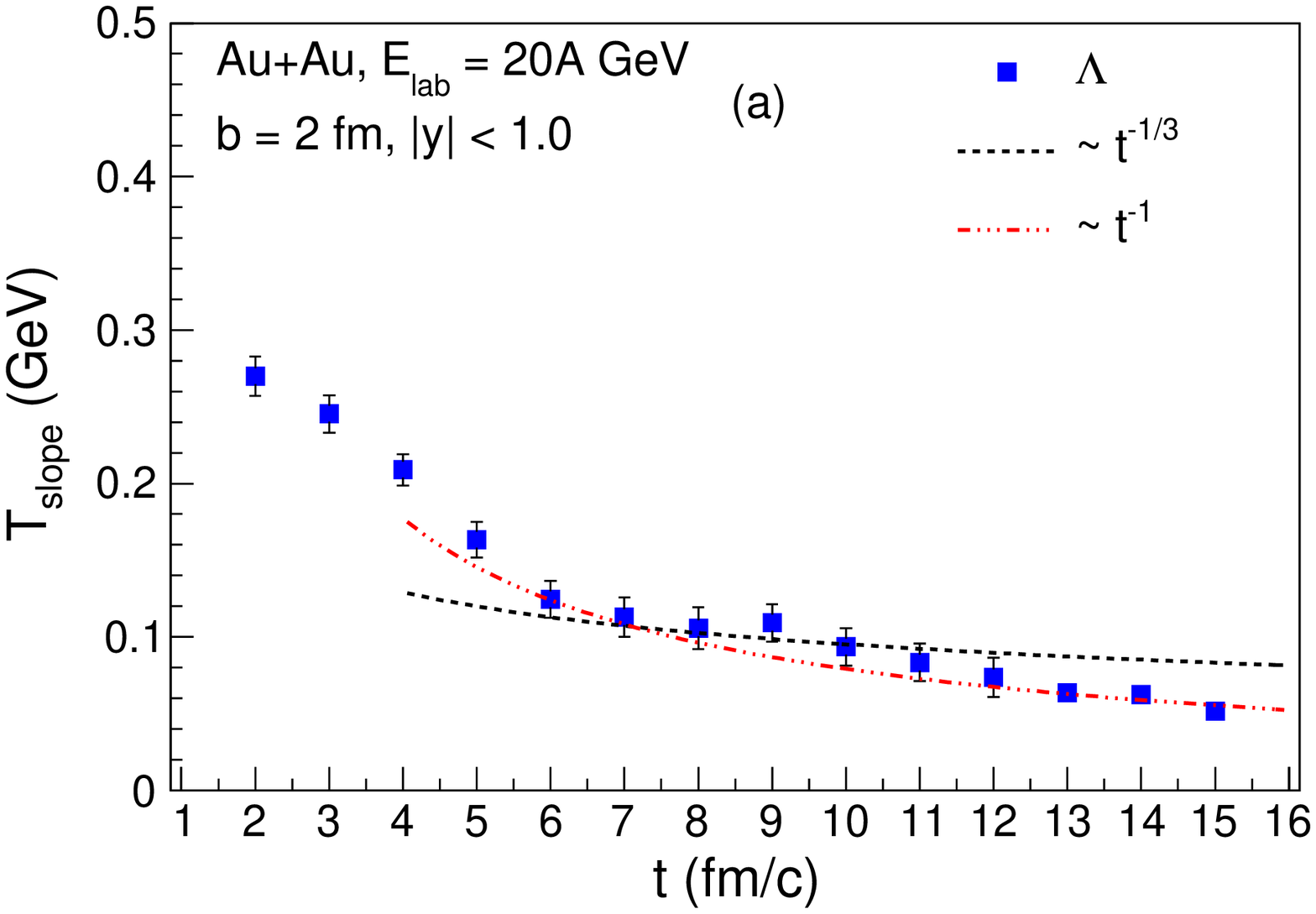}
\includegraphics[width= 7.5cm,clip= true]{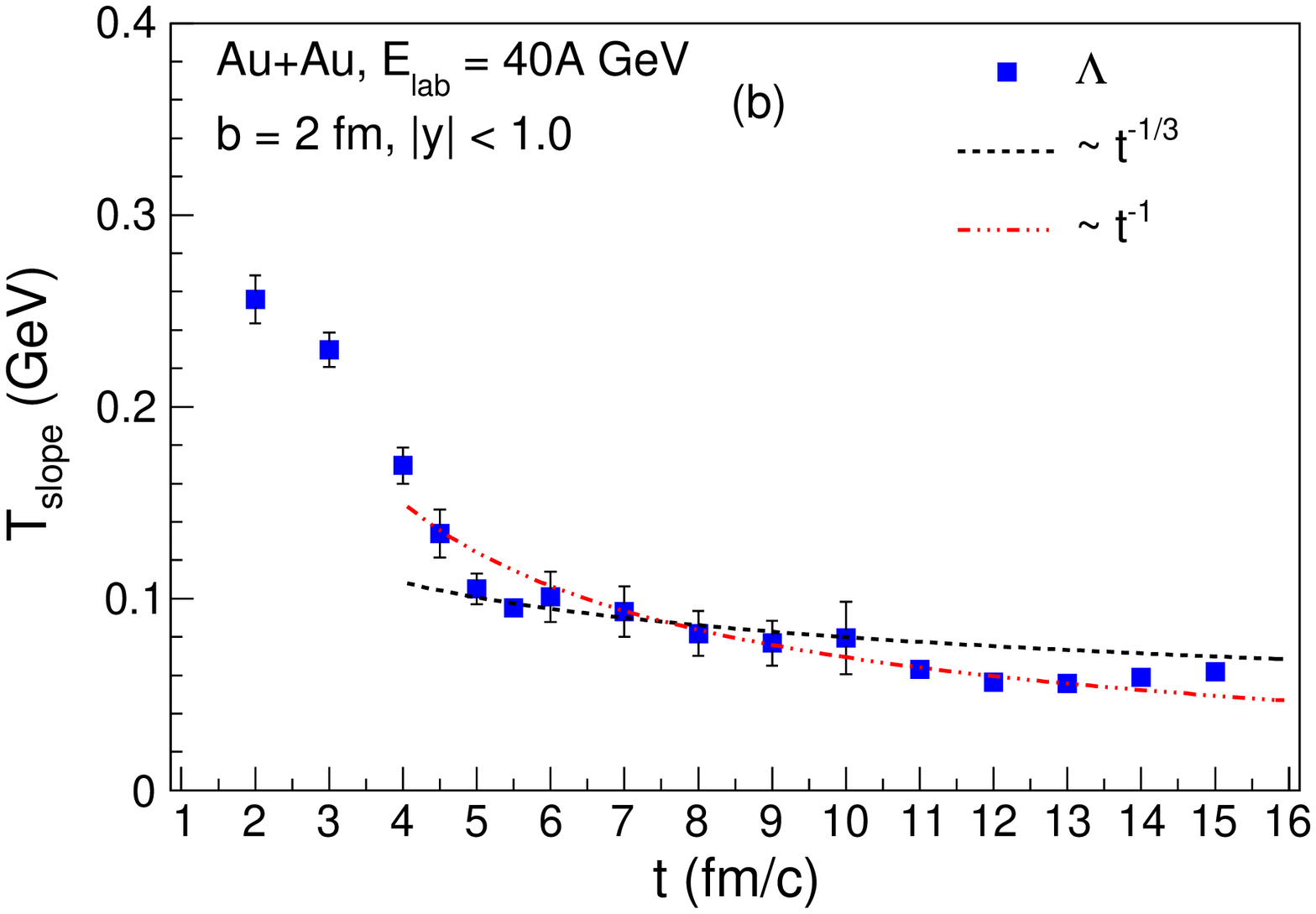}
\end{center}
\caption{(Color online) The scaling behaviour of inverse slope parameter ($T_{slope}$) of the energy spectra of Lambda 
inside the central cell for Au+Au collisions (b = 2 fm) at (a) 20A, (b) 40A GeV laboratory energies. The black dotted line 
denotes the scaling due to longitudinal expansion and the red dashed line denotes scaling due to three dimensional 
expansion. The error bars are statistical only.}
\label{fig-expansion}
\end{figure*}
\subsection{Thermalization of energy spectra of baryons}
In this section, we adopted an alternate approach to equilibrium which would reinforce the findings of earlier section. 
We investigated the time scale of local thermalization of baryonic matter from the time evolution of inverse slope parameter 
of the energy spectra ($EdN/d^3p$ vs E). For this purpose we have 
parameterized the energy spectra of Proton and Lambda inside the cell by Tsallis distribution~\cite{Tsallis}. An important 
criticism often arises that systems obeying non-extensive statistics achieve thermal equilibrium or not. Here we refer to the work of 
B\'{i}r\'{o} and Purcsel~\cite{TS.Biro} which has shown that two non-extensive subsystems do achieve a common equilibrium distribution 
within the framework of non-extensive Boltzmann equation. 
The Tsallis distribution has extensively been used in recent years for describing
the transverse momentum ($p_T$) distribution of produced hadrons at RHIC and the LHC energies~\cite{JCleymans, PShukla}. The special merit of 
the distribution is: at low energy limit it reduces to an exponential distribution and at high energy limit it reduces to a 
power-law distribution~\cite{C.Y.Wong}. Thus it can accommodate both equilibrium and non-equilibrium phenomena. 
A recent work has found that Tsallis distribution fits reasonably good all particle spectra for $p_T<$ 10 GeV at midrapidity in 
d~+Au, Cu~+~Cu, Au~+~Au collisions at RHIC~\cite{h.zheng}.
Keeping the facts in mind, 
we write the energy spectra of Proton and Lambda inside the cell of dimension $2\times2\times2$ fm$^3$ about the origin of 
Au~+~Au system as;
\be
E\frac{d^3N}{d^3p} = C(1+\frac{E}{bT})^{-b},
\ee
where E is the energy of baryon in the unit of GeV and b $=$ 1/(q-1) is dimensionless. C has the unit of Gev$^{-2}$ and T is in GeV. 
q is called the non-extensive parameter of Tsallis distribution. The values of C, b, T are obtained through fitting 
the energy spectra up to E $=$ 3 GeV. The inverse slope parameter of this distribution is given by:
\be
T_{slope} = T~+~(q-1)E.
\ee 
In the asymptotic limit $E\rightarrow0$, the inverse slope parameter ($T_{slope}$) gives the thermodynamic temperature of 
the system~\cite{TS.Biro}. We have 
calculated the $T_{slope}$ of proton and Lambda energy spectra at E $=$ 0.1 GeV (nearly pion mass) and studied its time evolution at the four 
beam energies. The error in $T_{slope}$ arises from the errors in the fitting parameters $T$ and $b$. 
The results are depicted in Fig.~\ref{fig-tslope}.

We have found $T_{slope}$ (for Proton and Lambda both) falls sharply with time 
and then almost scales as $\sim t^{-1/3}$ for t $\gtrsim$ 9 fm/c at 10A GeV laboratory energy. 
If $T_{slope}$ corresponds to the local temperature of system, then we can infer
an isentropic longitudinal expansion sets in inside the above mentioned cell analogous to Bjorken ideal hydrodynamics. 
The temperature follows the Bjorken scaling solution. We consider the time 
as the local thermal equilibration time scale of the system at which the scaling behaviour of 
the slope parameter has initiated. 
Similarly we have found the $t^{-1/3}$ scaling holds good for t$\gtrsim$ 7 fm/c at 20A GeV, 
t$\gtrsim$ 6 fm/c at 30A GeV and for t$\gtrsim$ 5 fm/c at 40A GeV beam energy.
At later times (say, t$>$ 10 fm/c at $\rm{E_{lab}}=$ 40A GeV), the $T_{slope}$ is seen 
to scale as $\sim t^{-1}$ owing to three dimensional 
spherical expansion of the system (see Fig.~\ref{fig-expansion}). The assumption of Bjorken hydrodynamic regime with the 
above mentioned scaling solution, namely, initial one-dimensional
flow and ideal gas equation of state, could be dubious at lower 
collision energies. Although earlier works at AGS and SPS energies~\cite{Kolb-AGS} have found phenomenological success 
based on it. However it may be noted that we do not study thermalization of whole reaction volume, rather concentrate at the very 
central part of the system only. And for this region the above assumptions may be relevant, at least we can
identify clearly the Bjorken scaling regime of $T_{slope}$ for all energies (see Fig.~\ref{fig-tslope}).
Thus we have found time scale of thermalization of energy spectra 
roughly in agrees with the pressure isotropization time of baryons and decreases with the 
increase in laboratory energy for the above mentioned cell.
\begin{figure}[tb]
\begin{center}
\includegraphics[width= 7.5cm, clip=true]{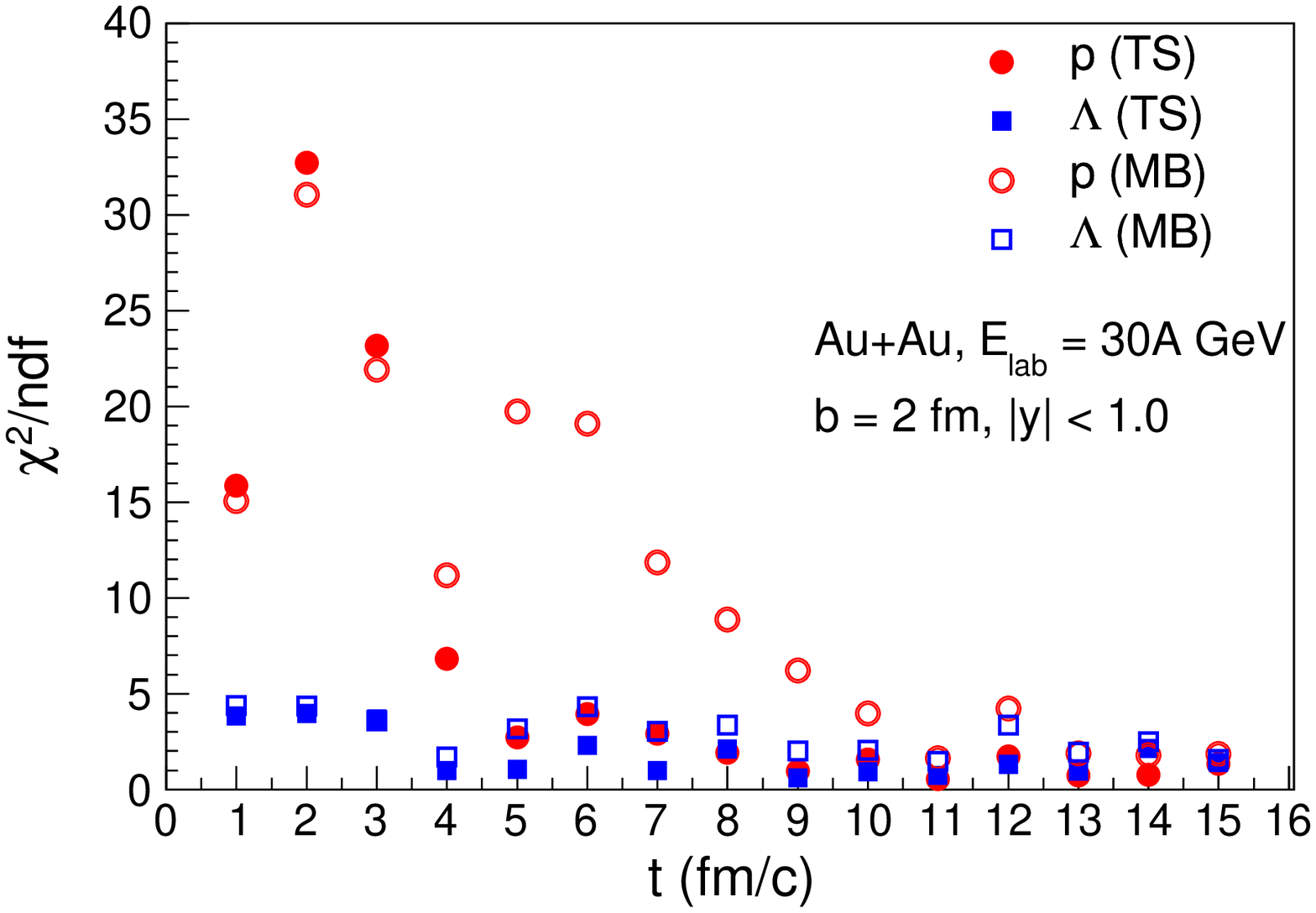}
\includegraphics[width= 7.5cm, clip=true]{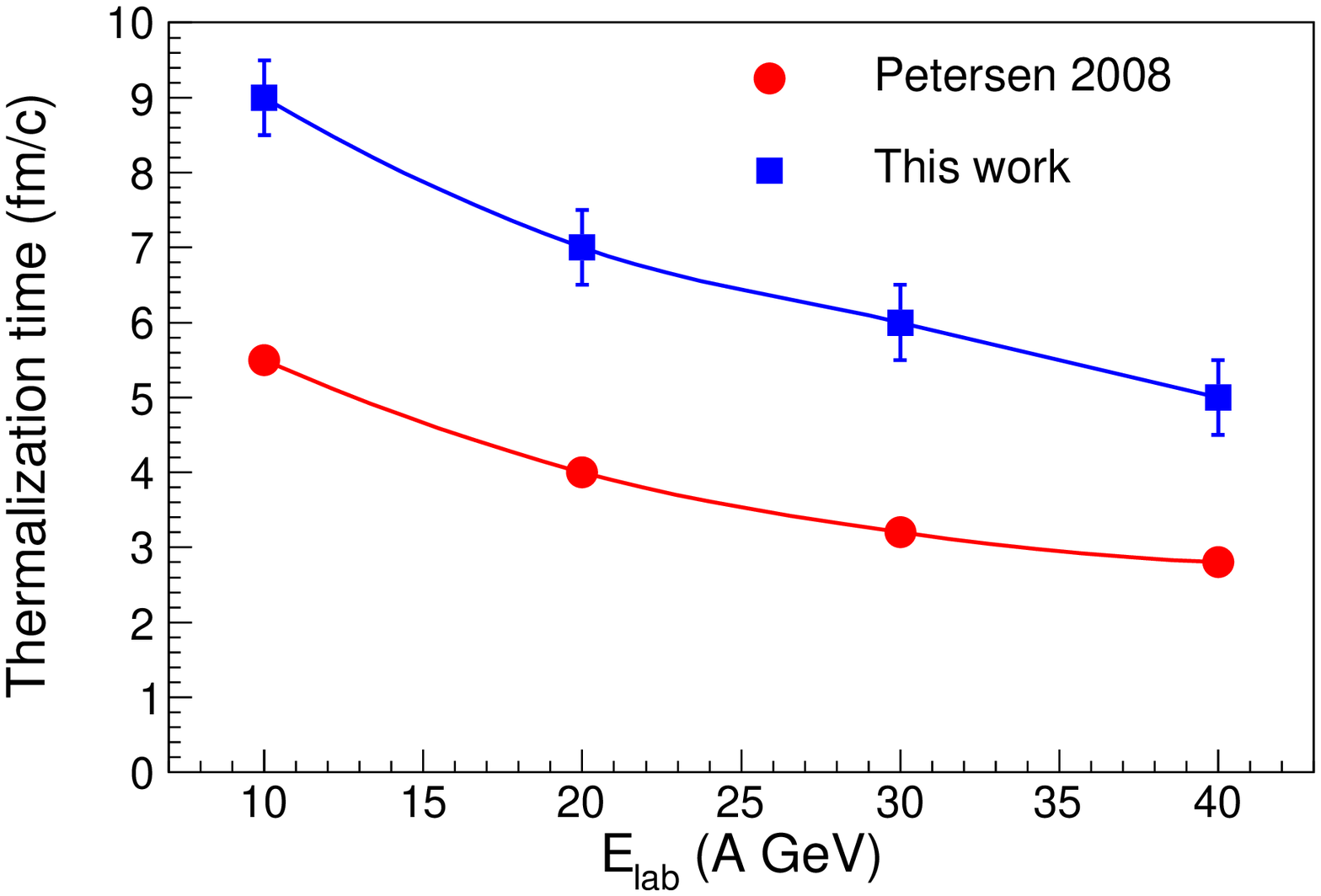}
\end{center}
\caption{(Color online) (Upper panel) The chi-square per degrees of freedom ($\chi^2/ ndf$) at different times for the 
Tsallis (filled symbols) and Maxwell-Boltzmann (open symbols) distribution which are fitted to the energy spectra of 
Proton and Lambda in the central cell for Au~+Au collisions (b$=$ 2 fm) at $\rm{E_{lab}}=$ 30A GeV. (Lower panel) The thermalization 
time obtained in this work (blue square) is compared with the local thermalization time $t_{start}$ (red circle) used in the 
hybrid (UrQMD~+~Hydrodynamics) model (Petersen 2008:\cite{H.Petersen}). The error bars are considered to be systematic.}
\label{fig-Tsallis}
\end{figure}

A natural question about the analysis involving $T_{slope}$ could arise that how better does the Tsallis distribution 
fit the spectrum compared to any classical distribution. In order to see that, we fit the energy spectra of baryons with 
Maxwell-Boltzmann (MB) distribution; $f(E) = C^{\prime} exp(-(E-\mu)/T)$. $C^{\prime}$ is a constant, $T$ is the 
temperature and $\mu$ is the chemical potential in usual notation. We have fitted the spectra for Au~+Au collisions at $\rm{E_{lab}}=$ 30A 
GeV for the same range of $E$ and calculated the chi-square per degrees of freedom ($\chi^2/ ndf$) for different times. 
The result is depicted in Fig.~\ref{fig-Tsallis}. It has been observed that Tsallis distribution mostly gives lower value of 
$\chi^2/ ndf$ which is close to unity in comparison to MB distribution. The inverse slope parameters of both distributions 
at different times are listed in Table~\ref{Table_tslope}. 

Several facts emerge upon close inspection. First, the two 
parameters are very similar at early times (say up to 3 fm/c). This might be due to numerical equivalance of the two distributions at 
these times. However it can be noted that $\chi^2/ ndf$ comes out very large at those times for both distributions; thus the 
parameters may not be describing a good fit.

Now in the thermal regime, say for t$\geq 6$ fm/c, the two parameters differ by nearly 40 MeV and $T_{slope}$ (Tsallis) is 
smaller than T(MB). The behaviour has been studied in the Ref.~\cite{Levai}. The Tsallis distribution describes a 
near-thermal equilibrium situation for $q$ value close to unity. For the same particle yield, Tsallis distribution leads to 
lower temperature (i.e. inverse slope parameter) than MB distribution for $q>$1.  The Tsallis temperatue often interpreted 
as the superposition different MB temperatures and the relaive width of fluctuation in T(MB) 
is related to non-extensivity parameter $(q-1)$~\cite{G.Wilk}. We have checked that $(q-1)$ remains almost 
constant$\sim0.03$ during the time span. The constant difference between the slope parameters can be attributed to this fact.
A similar trend between the inverse slope parameters has been reported in~\cite{Tsallis-MB} where the particle spectra for 
central Pb-Pb collisions at the LHC energy, are fitted with both Tsallis and MB distribution.
\begin{table*}[!]
\setlength{\arrayrulewidth}{0.25mm}
\setlength{\tabcolsep}{12pt}
\renewcommand{\arraystretch}{1.1}
\begin{tabular}{ccc} \hline
\multicolumn{1}{c}{t} & 
\multicolumn{1}{c}{$T_{slope}$(Tsallis)} &
\multicolumn{1}{c}{$T$(MB)}\\
fm/c & GeV & GeV \\
\hline\hline
1 & 0.275 & 0.267 \\ 
2 & 0.252 & 0.247 \\
4 & 0.200 & 0.223 \\
6 & 0.140 & 0.195 \\
8 & 0.138 & 0.178 \\
10 & 0.121 & 0.161 \\
12 & 0.099 & 0.138 \\
14 & 0.083 & 0.116 \\
\hline
\end{tabular}
\caption{The inverse slope parameters for Tsallis and Maxwell-Boltzmann distributions at different times. 
The distributions are fitted to the energy spectra of Protons in the central cell for Au+Au collisions at $\rm{E_{lab}}=$ 30A GeV.}
\label{Table_tslope}
\end{table*}
\subsection{A comparison with earlier work}
We have compared our result with the local thermalization time scale ($t_{start}$) used by an earlier work of 
hybrid model of Boltzmann transport and hydrodynamics by Petersen {\it et al.}~\cite{H.Petersen}. The model has successfully 
described the data of rapidity dependent yield, transverse mass spectra of hadrons at AGS and SPS experiments. 
The $t_{start}$ is considered ad hoc as the nuclear 
passage time in the center of mass frame. The comparison can be found in Fig.~\ref{fig-Tsallis}. We have introduced 
a systematic uncertainty of 
$\pm0.5$ fm in our estimated thermalization time because the simulation was carried out in time step $\Delta t=$ 1 fm. 
It has been found that our result decreases with increasing laboratory energy 
similar to $t_{start}$ but is about 1.5 times larger in magnitude. The earlier work
has assumed that $t_{start}$ is the lowest possible time needed for local thermalization 
however the current study could provide a more realistic estimate of it. Nevertheless, the issue has been 
investigated further in Ref.~\cite{H.Petersen} and found that multiplicity and mean transverse momenta of particles 
do not change appreciably when $t_{start}$ increases by factor of 2.
\begin{table*}
\setlength{\arrayrulewidth}{0.5mm}
\setlength{\tabcolsep}{12pt}
\renewcommand{\arraystretch}{1.25}
\begin{tabular}{|cccc|cccc|} \hline
\multicolumn{4}{|c|}{$E_{lab}$ = 10A GeV} & 
\multicolumn{4}{c|}{$E_{lab}$ = 30A GeV} \\
\hline t & T & $\mu_B$ & $\mu_s$ & t & T & $\mu_B$ & $\mu_s$ \\
fm/c & GeV & GeV & GeV & fm/c & GeV & GeV & GeV \\
\hline 10 & 0.145 & 0.708 & 0.174 & 8 & 0.152 & 0.616 & 0.123 \\
11 & 0.136 & 0.697 & 0.148 & 9 & 0.145 & 0.601 & 0.100 \\
12 & 0.128 & 0.687 & 0.125 & 10 & 0.137 & 0.595 & 0.081 \\
13 & 0.120 & 0.680 & 0.102 & 11 & 0.129 & 0.593 & 0.067 \\
14 & 0.114 & 0.670 & 0.082 & 12 & 0.123 & 0.587 & 0.047 \\
15 & 0.108 & 0.664 & 0.070 & 13 & 0.115 & 0.586 & 0.031 \\
16 & 0.102 & 0.659 & 0.049 & 14 & 0.110 & 0.586 & 0.019 \\
17 & 0.097 & 0.656 & 0.041 & 15 & 0.105 & 0.585 & 0.011 \\
\hline
\end{tabular}
\caption{The time evolution of temperature (T), baryon chemical potential($\mu_B$), strange chemical potential ($\mu_s$) in the central 
cell ($2\times2\times2$ fm$^3$) for Au~+~Au collisions (b = 2 fm) at laboratory energies 10A and 30A GeV. The thermodynamic parameters 
are obtained from energy density of baryons ($\varepsilon_B$), number density of baryons ($n_B$) and number density of
strange hadrons ($n_s$) using statistical hadron gas model.}
\label{Table_tmub}
\end{table*}
\section{Comparison with statistical thermal model}
\begin{figure}
\begin{center}
\includegraphics[width= 7.5cm, clip=true]{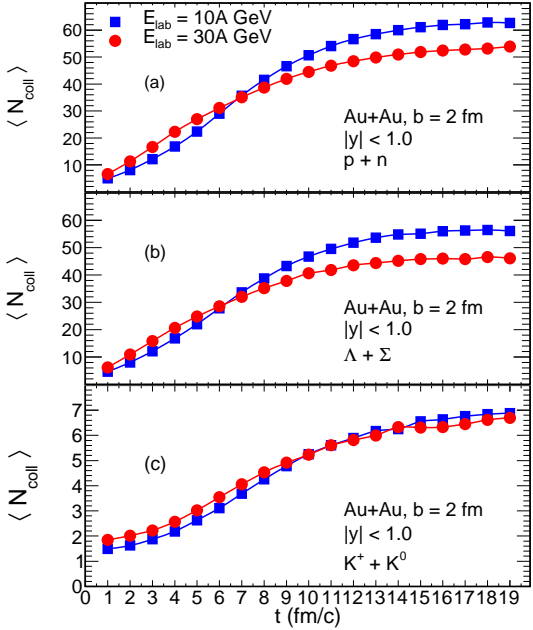}
\end{center}
\caption{(Color online) Time evolution of average number of collisions ($\langle N_{coll}(t)\rangle$) suffered by 
(a) Protons and Neutrons, (b) Lambda and Sigma baryons, (c) Kaons for Au~+~Au collisions (b = 2 fm) at laboratory 
energies 10A and 30A GeV.}
\label{Fig-Ncoll}
\end{figure}

In the preceding sections we argued that the dense hadronic matter created in the collisions will achieve local thermal equilibrium 
on a certain time scale. Thus we can employ the statistical hadron gas model~\cite{Stachel} 
to extract the intensive thermodynamic variables like; temperature, chemical potential of the system during subsequent
evolution.
The statistical model can not be applied prior to equilibrium, rather can be applied beyond thermal freeze-out 
of the system. Traditionally thermal freeze-out is defined as: the average 
scattering rate between the constituents becomes smaller than the average expansion rate of the system. The system has become so dilute 
that hardly any collision between the constituents takes place. Following this criterion, we have checked the time evolution of
the average number of collisions ($\langle N_{coll}(t)\rangle$) suffered by different hadron species. The 
Fig.~\ref{Fig-Ncoll} shows that average number of collisions suffered by p, n, $\Lambda$, $\Sigma$ baryons and $K$ mesons almost 
saturate for  t$\gtrsim$ 17 fm/c at $\rm{E_{lab}}=$ 10A GeV and t $\gtrsim$ 15 fm/c at $\rm{E_{lab}}=$ 30A GeV.
Considering the above scenarios; we made the comparison of statistical model with UrQMD during the  
time interval 10 fm/c $\leq t\leq$ 17 fm/c at 10A GeV and 8 fm/c $\leq t\leq$ 15 fm/c at 30A GeV laboratory energy. 

\begin{figure}
\begin{center}
\includegraphics[width= 8.0cm, clip=true]{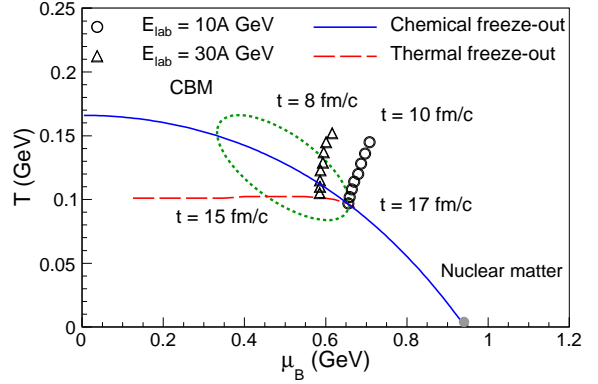}
\end{center}
\caption{(Color online) The evolution of temperature ($T$) and baryo-chemical potential ($\mu_B$) in the central cell for 
Au~+~Au collisions (b = 2 fm) at $\rm{E_{lab}}$ = 10A GeV and 30A GeV.The solid (blue) line 
denotes the chemical freeze-out and the dashed (red) line denotes thermal freeze-out boundary in relativistic 
heavy ion collisions~\cite{chem-freeze,uli-hydro}. The dotted circle denotes the expected region probed
by the CBM experiment ($\sqrt{s_{NN}}$ = 4--10 GeV) at FAIR.}
\label{Fig-Tmub}
\end{figure}

The expression for number density, energy density  for the i' th hadron species in the statistical hadron gas model 
are given by:
\bea
n_i &=& \frac{g_i}{(2\pi\hbar)^3}\int 4\pi p^2 f_i(T,\mu_i)\,dp, \nn \\
\varepsilon_i &=& \frac{g_i}{(2\pi\hbar)^3}\int 4\pi p^2 e_i\,f_i(T,\mu_i)\,dp, \nn
\eea
where $e_i$ is the energy, $T$ is the temperature  and $\mu_i$ is chemical potential of the i'th hadron. 
The hadrons are considered relativistic, $e_i=(p^2+m_i^2)^{\frac{1}{2}}$. $f_i$ is the distribution function of 
the i'th hadron (either Fermi-Dirac or Bose-Einstein).
However above distributions are practically approximated to classical 
MB distribution as; ($e_i-\mu_i$)/T$>>$1. $\mu_i$ can be decomposed in terms of 
baryonic ($\mu_B$) and strange ($\mu_s$) chemical potentials. The charge chemical potential ($\mu_Q$), 
which is an order of magnitude smaller than the other two, has been neglected here.
\be
\mu_i = b_i\mu_B~+~s_i\mu_s,   \nn
\ee
$b$ and $s$ are the baryon and strangeness quantum number respectively. $T$, $\mu_B$, and $\mu_s$ are extracted from 
the following equations;
\be
\varepsilon_B = \sum_i^{baryon}\varepsilon_i,\hspace*{0.15cm}n_B = \sum_i^{baryon}b_in_i,\hspace{0.15cm}n_s = \sum_i^{baryon,meson}s_in_i
\label{tmub}
\ee
The quantities in the l.h.s. of the equation~\ref{tmub}, namely energy density of baryons ($\varepsilon_B$), 
number density of baryons ($n_B$) and number density of strange hadrons ($n_s$) are obtained from the UrQMD. We have solved 
the above set of equations during the time interval stated earlier. The values are listed in Table~\ref{Table_tmub}. We have 
plotted them in the QCD phase diagram in order to get an estimate about the chemical and thermal 
freeze-out time of the system (see Fig.~\ref{Fig-Tmub}).
The chemical freeze-out line has been obtained empirically from the thermal model fit of particle ratios at different 
collision energies~\cite{chem-freeze}. The thermal or kinetic freeze-out line has also been obtained phenomenologically 
from the blast wave model fits of the measured hadron spectra at different experiments~\cite{uli-hydro}. It can be seen 
at low energies $\rm{E_{lab}}$ = 10A GeV, the chemical and the kinetic freeze-out happens almost instantaneously 
at t$\approx$17 fm/c. 
At higher energy $\rm{E_{lab}}$ = 30A GeV, system undergoes first chemical freeze-out at t$\approx$13 fm/c, then 
kinetic freeze-out at t$\approx$15 fm/c. The feature has already observed in low energy collision 
experiments at RHIC~\cite{Phase-diagram}. 
We would also like to add that our estimation of temperatures at the kinetic 
freeze-out times closely agree with the values given by blast wave model fit to the $\Lambda$ baryon spectra from 
NA49 Collaboration at the similar laboratory energies~\cite{NA49}.

\begin{figure}
\begin{center}
\includegraphics[width= 8.0cm, clip=true]{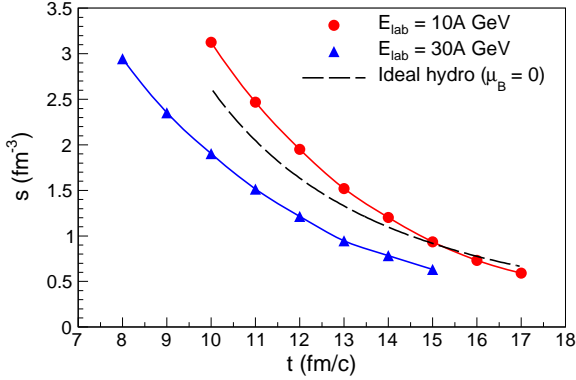}
\end{center}
\caption{(Color online) The time evolution of entropy density of baryonic matter inside the central cell for 
Au~+Au collisions (b = 2 fm) at $\rm{E_{lab}}$ = 10A GeV and 30A GeV. The dashed line denotes the parameterization 
of ideal hydrodynamic evolution of entropy density in the central region for Au~+Au collisions 
(b = 0 fm) at $\sqrt{s_{NN}}$ = 200 GeV~\cite{P.Kolb}.}
\label{Fig-Entropy}
\end{figure}

We are interested in computing bulk properties of a baryon rich hadronic medium, thus strange meson contribution can be neglected 
as $\mu_s\rho_M^{S}\approx$ few MeV.
Using the values of temperature and chemical potential listed in Table~\ref{Table_tmub}, we have calculated the pressure of 
baryons with the statistical hadron gas model:
\be
P = \sum_i^{baryon} \frac{g_i}{(2\pi\hbar)^3}\int 4\pi p^2dp \frac{p^2}{3(p^2+m_i^2)^{\frac{1}{2}}} f_i(T,\mu_i),
\ee
and the entropy density ($s$) for baryons using the thermodynamic relation:
\be
Ts = \varepsilon_B~+~P~-\mu_i(\rho_B^{NS}~+~\rho_B^{S}),
\ee
where $\mu_i$ is the chemical potential, defined earlier in this section.
We studied the time evolution of entropy density at $\rm{E_{lab}}$ = 10A and 30A GeV till the thermal decoupling. 
Our aim is to get some insight about the fluidity of the dense baryonic matter created in these collisions. In recent 
times several calculations~\cite{Itakura,Denicol,H.Mishra} have been reported on the transport properties of hadronic matter at 
finite baryo-chemical potential, including the effect high mass resonances, etc. However the fluidity of 
dense hadronic matter has possibly first discussed in~\cite{Denicol} and subsequently in~\cite{S.Ghosh,G.Kadam}. 
The authors of Ref.~\cite{Denicol} have argued that the fluid behaviour of a baryon rich 
($\mu_B\sim$ 500 MeV) hadron gas is closer to the ideal fluid limit than the corresponding gas with zero baryon number. Following their 
observation, we have compared the entropy densities at $\rm{E_{lab}}$ = 10A and 30A GeV with the ideal fluid limit 
reached at the highest RHIC energy ($\sqrt{s_{NN}}$ = 200 GeV). We have 
parameterized temporal evolution of entropy density of hadronic matter from an ideal hydrodynamic simulation~\cite{P.Kolb} 
for central Au~+Au collisions at $\sqrt{s_{NN}}$ = 200 GeV. The entropy density at r = 3 fm from the center has 
been found to scale with 
proper time ($\tau$) as$\sim \tau^{-2.6}$ for $\tau\geq$ 10 fm/c. The results are depicted in Fig.~\ref{Fig-Entropy} along with the 
parameterization from ideal hydrodynamics.
It is heartening to see that the evolution of entropy density at $\rm{E_{lab}}$ = 30A GeV closely resembles with 
ideal hydrodynamic limit at zero net baryon density. The entropy density at $\rm{E_{lab}}$ = 10A GeV falls even little faster than the 
aforementioned limit. It may imply that the hadronic matter produced at 10A GeV beam energy is more ideal 
than the same at 30A GeV beam energy. The observation can be understood using the fact that 
shear viscosity to entropy density ratio ($\eta/s$) of a hadronic system decreases with increasing fugacity ($\mu_B/T$) of 
the system~\cite{Bass-viscosity}.
\section{Summary and Discussion}
In this article, we have investigated the time scale for local thermal equilibration of dense baryonic matter created in 
central Au~+~Au collisions at the proposed CBM experiment energies of $\rm{E_{lab}}$ = 10A, 20A, 30A, 40A GeV. The microscopic transport 
model UrQMD has been used for this purpose in the default cascade mode. The net baryon density has been found maximum at 30-40 GeV 
and the net strangeness of the created hadronic matter is dominated by baryons for all energies stated above. We have studied the 
time evolution of longitudinal-to-transverse microscopic pressure anisotropy and inverse slope parameter of the energy spectra of baryons 
and mesons inside a cell of 8 fm$^3$ in the central region of Au~+~Au system. The pressure anisotropy ratio of baryons and mesons has achieved a constant value close to unity, on a certain time. 
The time has been found to decrease with the increase in laboratory energy. The time scale obtained from the evolution of 
inverse slope parameter of energy spectra of baryons nearly agrees with the pressure (or momentum) isotropization time. 
However a small time difference ($\Delta t\sim$ 0.5 fm/c) in the pressure isotropization as well as in the 
thermalization of energy spectra between strange and non-strange baryons has been noticed.
We have chosen our test volume in the central collision zone. The estimated time scales are expected 
to grow in a region which is away from the center because of small scattering rates. Therefore the present study 
provides a realistic estimate of the time scales required for achieving thermodynamic equilibrium  for the 
central region of the system created at those energies.

Using the statistical thermal model, we have obtained the temperature and chemical potentials of the hadronic matter during the 
post-equilibrium evolution at $\rm{E_{lab}}$ = 10A and 30A GeV. They are found to agree qualitatively with the empirical relation between 
T and $\mu_B$ at the chemical freeze-out. In addition we have calculated the entropy density of the baryonic matter inside the cell and 
found the evolution is quasi-isentropic, close to the ideal hydrodynamic limit at zero net baryon density.
\section{Acknowledgement}
The work is partially supported by Bose Institute Indo-Fair Coordination Centre (BI-IFCC).
We are thankful to Grid tier-2 center, Kolkata for providing computing resources. We acknowledge the financial assistance from 
DAE, India and FAPESP, Brasil, during the course of the work. 

\end{document}